\documentclass[aps,prd,groupedaddress,preprintnumbers,
              floatfix, nofootinbib,longbibliography]{revtex4}
\usepackage[utf8]{inputenc}
\usepackage{hyperref}
\usepackage{slashed}
\usepackage{natbib}
\usepackage{graphicx}
\usepackage{epsfig}
\usepackage{amsmath}
\input{epsf}
\usepackage{psfrag}

\usepackage[usenames,dvipsnames]{color}

\newcommand{\beq}{\begin{eqnarray}}
\newcommand{\eeq}{\end{eqnarray}}
\newcommand{\Slash}[1]{{\ooalign{\hfil/\hfil\crcr$#1$}}}

\newcommand{\nn}{\nonumber \\}

\usepackage{amsmath,color}
\usepackage{amssymb}
\usepackage{bm}
\usepackage{graphicx}
\usepackage{multirow}

\usepackage{soul} 
\newcounter{RSQ}

\begin{document}
\preprint{LA-UR-24-31861}
\title{Nonlocal chiral anomaly and generalized parton distributions}

\author{Shohini Bhattacharya}
\affiliation{Theoretical Division, Los Alamos National Laboratory, Los Alamos, New Mexico 87545, USA}

\author{Yoshitaka Hatta}
\affiliation{Physics Department, Brookhaven National Laboratory, Upton, NY 11973, USA}
\affiliation{RIKEN BNL Research Center, Brookhaven National Laboratory, Upton, NY 11973, USA}

\author{Jakob Schoenleber}
\affiliation{RIKEN BNL Research Center, Brookhaven National Laboratory, Upton, NY 11973, USA}

\begin{abstract}
We discuss the nonlocal generalization of the QCD chiral anomaly along the light-cone and derive relations between twist-two, twist-three and twist-four generalized parton distributions (GPDs) mediated by the  anomaly. We 
further establish the connection to the `anomaly pole' in the GPD $\tilde{E}$ recently identified in the perturbative calculation of the Compton scattering amplitudes, and demonstrate its cancellation at the GPD level. Our work helps elucidate the previously unexplored connection between GPDs, the  chiral anomaly, and the mass generation of the $\eta'$ meson. 
\vspace*{0.4cm}
\noindent

\end{abstract}

\maketitle

\section{Introduction}

 In previous work \cite{Bhattacharya:2022xxw, Bhattacharya:2023wvy}, two of us examined the potential manifestations of QCD anomalies,  specifically the chiral and trace anomalies, in Deeply Virtual Compton Scattering (DVCS), aiming to uncover novel insights into Generalized Parton Distributions (GPDs). This is along the line of the earlier works \cite{Jaffe:1989jz,Tarasov:2020cwl,Tarasov:2021yll} on the connection between the chiral anomaly and  polarized Deep Inelastic Scattering (DIS) through the `anomaly pole'. 
It is simple to explain why QCD anomalies have something to do with GPDs.  Consider the chiral anomaly in the divergence of the singlet axial current in massless QCD with $n_f$ flavors \cite{Adler:1969gk,Bell:1969ts}
\beq
\partial_\mu J^\mu_5 = -\frac{n_f\alpha_s}{4\pi}F^{\mu\nu}\tilde{F}_{\mu\nu}. \label{ad} 
\eeq
The off-forward proton matrix element of this relation is 
\beq
g_A(t)+\frac{t}{4M^2}g_P(t) = \frac{i}{2M}\frac{\langle p'|\frac{n_f\alpha_s}{4\pi}F\tilde{F}|p\rangle}{\bar{u}(p')\gamma_5u(p)} , \label{form1}
\eeq
 where $g_A(t)$ and $g_P(t)$ with $t=(p'-p)^2$ being the momentum transfer are the standard proton axial form factors and $M$ is the proton mass. 
 (\ref{form1}) clearly shows that (i) the axial form factors are related by the chiral anomaly. We then recall that (ii) the axial form factors are moments of the polarized quark GPDs $\tilde{H}$ and $\tilde{E}$ \cite{Ji:1998pc}
 \beq
 g_A(t)=\int dx \tilde{H}(x,t), \qquad g_P(t)=\int dx \tilde{E}(x, t).
 \eeq
 An immediate logical consequence of these two facts is that  the polarized GPDs $\tilde{H}$ and $\tilde{E}$ are related by the chiral anomaly. 
Essentially the same argument can be made between the unpolarized GPDs, the gravitational form factors, and the trace anomaly \cite{Bhattacharya:2022xxw,Bhattacharya:2023wvy}.  

In pursuit of such relations, the authors of  
\cite{Bhattacharya:2022xxw,Bhattacharya:2023wvy} have performed the one-loop calculation of DVCS amplitudes using $t$ as the collinear regulator,  following a similar calculation for polarized DIS \cite{Tarasov:2020cwl,Tarasov:2021yll}. It has been shown that the chiral and trace anomalies  manifest themselves as infrared  poles  $1/t$ in the amplitudes, and these poles can be systematically absorbed into the corresponding GPDs  via the standard infrared subtraction procedure  which is built in into a proper factorization formalism.  
This means that,  for example,   the GPD $\tilde{E}$ contains a term proportional to the chiral anomaly pole    
\beq
\tilde{E}(x,\xi,t) = \frac{4M^2}{t} \frac{n_f\alpha_s}{2\pi} C\otimes \tilde{\cal F}(x,\xi,t)+\cdots,
\label{et}
\eeq
where $C$ is a certain convolution kernel and $\tilde{\cal F}$ is a twist-four GPD  
\beq
\tilde{{\cal F}}(x,\xi,t) \propto \int \frac{dz^-}{2\pi} e^{ixP^+z^-} \langle p'|F^{\mu\nu}(-z^-/2)W\tilde{F}_{\mu\nu}(z^-/2)|p\rangle,  \label{4}
\eeq
which features a nonlocal generalization of the local operator $F^{\mu\nu}\tilde{F}_{\mu\nu}$ \cite{Mueller:1997zu,Agaev:2014wna,Tarasov:2020cwl,Hatta:2020ltd,Radyushkin:2022qvt,Bhattacharya:2022xxw}. Of course, there is no massless pole in $\tilde{E}$ in QCD (even in the chiral limit), so the pole at $t=0$ must be canceled by another massless pole due to the exchange of the `primordial' $\eta_0$ meson, the would-be Nambu-Goldstone boson of the broken U$_A$(1) symmetry. After this pole cancellation, the pole of $\tilde{E}$ is shifted to the physical $\eta'$ mass $t=m_{\eta'}^2\approx (957\, {\rm MeV})^2$. This is consistent with the well-known scenario of how the $\eta'$ meson  acquires mass due to the chiral anomaly \cite{Witten:1979vv,Veneziano:1979ec}. 
 While the implications of this phenomenon in polarized DIS, particularly the nonlocal relation between the $g_1$ structure function and the chiral anomaly, have been recently discussed \cite{Tarasov:2020cwl,Tarasov:2021yll},  they have not been fully explored in the context of GPDs \cite{Bhattacharya:2022xxw,Bhattacharya:2023wvy} (see also \cite{Bass:2001dg}).

 A drawback of the argument in \cite{Bhattacharya:2022xxw,Bhattacharya:2023wvy} is that a ‘counter GPD’ of nonperturbative origin had to be introduced by hand to cancel the $1/t$ pole in (\ref{et}). Moreover, $1/t$ poles have been identified in partonic scattering amplitudes and GPDs evaluated using partonic matrix elements. It has been suggested that they should be ‘promoted’ to hadronic (proton) matrix elements to maintain consistency with form factor relations such as (\ref{form1}), although this has not been explicitly demonstrated. In this paper, we address both of these issues by deriving nonperturbative relations between the twist-two, twist-three, and twist-four GPDs of the proton mediated by the chiral anomaly. We then confirm the anomaly pole in (\ref{et}) and determine the missing terms in this equation.

\section{Nonlocal chiral anomaly}

The chiral anomaly relation (\ref{ad}) more precisely reads 
\beq
\partial_\mu J^\mu_5(x) =\partial_\mu (\bar{\psi}\gamma^\mu \gamma_5 \psi )= 2im_q\bar{\psi}\gamma_5\psi -\frac{n_f\alpha_s}{4\pi}F_a^{\mu\nu}\tilde{F}^a_{\mu\nu} ,\label{ad2}
\eeq
where $a=1,2,\cdots,8$ and $\tilde{F}^{\mu \nu} = \tfrac{1}{2} \epsilon^{\mu \nu \rho \sigma}F_{\rho \sigma}$.  The summation over $n_f$ degenerate quark flavors is implicit. We mostly assume massless QCD $m_q=0$ and comment on the effect of the current quark mass toward the end.  
Our convention is the same as in \cite{Bhattacharya:2023wvy}, namely,   
$\gamma_5=i\gamma^0\gamma^1\gamma^2\gamma^3$ and $\epsilon^{0123}=-\epsilon_{0123}=+1$ so that ${\rm Tr}[\gamma^\mu \gamma^\nu \gamma^\rho \gamma^\lambda \gamma_5]=-4i\epsilon^{\mu\nu\rho\lambda}$. The covariant derivative reads  $D_\mu = \partial_\mu+igA_\mu$. The light-cone coordinates are defined as $z^\pm =\frac{1}{\sqrt{2}}(z^0\pm z^3)$ and the transverse components $z^i$ are indicated by Latin indices $i,j=1,2$.  Among the various derivations of (\ref{ad2}), the most relevant for us is the point splitting method \cite{Peskin:1995ev} in which the current is slightly made non-local
\beq
J_5^\mu(x)=\lim_{z\to 0} \bar{\psi}\left(x-\frac{z}{2}\right)\gamma^\mu \gamma_5 P\exp\left(ig\int^{x+\frac{z}{2}}_{x-\frac{z}{2}}dw_\nu A^\nu(w)\right) \psi\left(x+\frac{z}{2}\right),
\eeq
where a straight Wilson has been inserted to maintain gauge invariance. 
As usual, the divergence of the current naively vanishes as $z^\mu\to 0$, but one finds a singularity  $\sim 1/z^2$ in the short distance behavior of the operator product $\psi(x+\frac{z}{2})\bar{\psi}(x-\frac{z}{2})$.  
After taking the limit $|z^\mu|\to 0$ in a  symmetrical way
\beq
\lim_{\epsilon\to 0}\frac{z^\nu z^\rho}{z^2}\to \frac{g^{\nu\rho}}{4} ,\label{limit}
\eeq
one recovers the anomaly term. 

In  \cite{Mueller:1997zu}, M\"uller and Teryaev have observed that if $z^\mu$ is taken along the light-cone $z^2= 0$, say in the minus direction $z^\mu = \delta^\mu_- z^-$, an anomaly still arises  from a UV divergence even if $z^-$ is kept finite. A closely related  observation has been made in   \cite{Agaev:2014wna} (see Appendix (A2) there) in a different context. 
In the following, we shall revisit these studies and present a refined derivation of the nonlocal chiral anomaly. Our starting point is the  operator identity  
\beq
 {\cal D}_\mu \left[\bar{\psi}(z_2)\gamma^\mu \gamma_5 W_{z_2,z_1}\psi(z_1)\right] &=&2im_q\bar{\psi}(z_2)\gamma_5 W\psi(z_1) +O_F(z_2,z_1) \nn 
 &&+ \bar{\psi}(z_2) \left[W(\Slash D-im_q)\gamma_5 +(\overleftarrow{\Slash D}-im_q)\gamma_5W\right]\psi(z_1)
 ,\label{nonlocal} 
 \eeq 
where 
 \beq
 O_F(z_2,z_1)\equiv iz^\nu\int_{0}^{1} d\alpha \bar{\psi}(z_2)\gamma^\mu \gamma_5  WgF_{\mu\nu}(z_{21}^\alpha)W \psi(z_1) , \label{of}
\eeq
with 
\beq
z^\mu\equiv z^\mu_1-z^\mu_2   , \qquad 
(z_{21}^\alpha)^{ \mu} \equiv \alpha z^\mu_2+(1-\alpha)z^\mu_1. 
\eeq
We initially assume that $z^\mu$ is space-like, $z^2<0$. 
$W$ is the straight Wilson line in the fundamental representation. We often  suppress the subscript on $W$  indicating the endpoints of the straight line, but they should be obvious by inspection. 
  ${\cal D}_\mu$ is the translation operator that acts on  any product of operators as  
\beq
{\cal D}_\mu [A(x)\cdots B(y)]  = \lim_{\epsilon^\mu\to 0} \frac{1}{\epsilon^\mu}\left[A(x+\epsilon)\cdots B(y+\epsilon) - A(x)\cdots B(y)\right].
\eeq
The second line of (\ref{nonlocal}) naively  vanishes owing to the equation of motion. However, a UV divergence arises in the limit $z^2\to 0$, and certain regularization schemes violate the  anti-commutativity between $\gamma_5$ and $\Slash D$. This generates an extra term which needs to be removed by a finite renormalization of the axial current operator. Since this issue is  understood in the literature \cite{Collins:1984xc,Mueller:1997zu}, it will not be discussed here.

In the limit $z^\mu\to 0$ (including $z^-\to 0$), the left hand side of (\ref{nonlocal}) reduces to the divergence $\partial_\mu J_{5}^\mu$. Thus the anomaly must be hidden in $O_F$ despite its explicit proportionality to $z^\nu$. 
Let us write it as 
\beq
O_F(z_2,z_1)=-iz^\nu \int_{0}^{1} d\alpha
{\rm Tr}\Bigl[\psi(z_1)\bar{\psi}(z_2) \gamma^\mu \gamma_5WgF_{\mu\nu}(z_{21}^\alpha)W \Bigr].
\label{subst}
\eeq 
When $z=z_1-z_2$ approaches the light-cone $z^\mu \to \delta^\mu_-z^-$, the quark bilinear $\psi(z_1)\bar{\psi}(z_2)$ develops  singularities which can be expanded in $1/z^2$ in $d=4-2\epsilon$ dimensions 
\cite{Balitsky:1983sw,Balitsky:1987bk}
\beq
&&\psi(z_1)\bar{\psi}(z_2) = \frac{i\Slash z}{2\pi^2 (z^2)^2}W_{z_1,z_2}-\frac{iz^\rho}{8\pi^2 z^2}\int_{0}^{1} d\beta W_{z_1,z_{12}^\beta}g\tilde{F}_{\rho\lambda}(z_{12}^\beta)W_{z_{12}^\beta,z_2}\gamma^\lambda \gamma_5  \label{exact} \\ 
&&\qquad  +\frac{i}{32\pi^2}\left(\frac{1}{\epsilon_{IR}}+
\ln\frac{-z^2\mu_{IR}^2e^{2\gamma_E}}{4}\right)
\Biggl[ g\int_0^1d\alpha \alpha(1-\alpha)  z_\mu D^2\tilde{F}^{\mu\nu}(z_{12}^\alpha)\gamma_\nu\gamma_5   +ig^2z_\mu\int_{0}^{1} d\alpha \int_{0}^\alpha d\beta \nn 
&& \qquad \qquad\times \left\{ (1-2\alpha+2\beta) F^{\mu\lambda}(z_{12}^\alpha)\tilde{F}_{\lambda \rho}(z_{12}^\beta)\gamma^\rho  +  \tilde{F}^{\mu\lambda}(z_{12}^\alpha)F_{\lambda \rho}(z_{12}^\beta) \gamma^\rho   +\beta F^{\rho\nu}(z_{12}^\alpha)\tilde{F}_{\rho\nu}(z_{12}^\beta)\gamma^\mu \right\}\gamma_5+\cdots   \Biggr] + {\cal O}(z^2), \nonumber
\eeq
where we neglected the quark mass  and $\mu^2_{IR}$ is the $\overline{\rm MS}$ scheme scale parameter associated with the infrared  divergence.   Apart from the leading term, we have  kept only the terms that contain a $\gamma_5$ so that they survive when inserted into (\ref{subst}). For simplicity, we omitted Wilson lines in the logarithmic terms $\ln z^2$. When substituted into (\ref{subst}), these terms  constitute the renormalization group evolution kernel of the twist-four operator $O_F\sim \bar{\psi}F^{+\mu}\gamma_\mu\gamma_5 \psi$.   (\ref{exact})  only includes the $g\to q$ splitting kernel of the evolution exhibiting the mixing with three-gluon, twist-four operators such as $F^{+\mu}F^{+\lambda}\tilde{F}_{\mu\lambda}$. In principle, at this order one has to  include the complete evolution kernel including also the $q\to q$ kernel and other contributions  \cite{Balitsky:1987bk,Braun:2009vc}. 

Let us focus on  the $1/z^2$ term. We substitute it into (\ref{subst}) and find
\beq
&&\frac{n_f g^2}{4\pi^2z^2}\int_{0}^{1} d\alpha \int_{0}^{1} d\beta  {\rm Tr}\left[\tilde{F}_{\rho\mu}(z_{12}^\beta)WF^{\mu\nu}(z_{21}^\alpha)W+F_{\rho\mu}(z_{12}^\beta) W\tilde{F}^{\mu\nu}(z_{21}^\alpha)W\right]z^\rho z_\nu \nn 
&&= -\frac{n_f\alpha_s}{2\pi}\int_{0}^{1} d\alpha \int_{0}^{1} d\beta  {\rm Tr}\left[F^{\mu\nu}(z_{12}^\beta) W\tilde{F}_{\mu\nu}(z_{21}^\alpha)W \right] \nn 
&& =-\frac{n_f\alpha_s}{4\pi}\int_{0}^{1} d\alpha \int_{0}^{1} d\beta  F^{\mu\nu}(z_{12}^\beta) \tilde{W}\tilde{F}_{\mu\nu}(z_{21}^\alpha),
\nn &&= -\frac{n_f\alpha_s}{2\pi}\int_{0}^{1} d\alpha \int_{0}^{1-\alpha} d\beta  F^{\mu\nu}(z_{12}^\beta) \tilde{W}\tilde{F}_{\mu\nu}\left(z_{21}^\alpha\right),
\label{anom}
\eeq
where $\tilde{W}$ is the Wilson line in the adjoint representation. In the first line, we have used the cyclicity of the trace and the symmetry of the $\alpha,\beta$-integrals. This allows us to utilize the formula\footnote{ A quick derivation goes as follows  
\beq
z^\nu F_{\nu\mu}W\tilde{F}^{\mu\rho}z_\rho=\frac{1}{2}z^\nu F_{\nu\mu}W\epsilon^{\mu\rho\alpha\beta}F_{\alpha\beta}z_\rho &=& \frac{1}{2}F_{\nu\mu}WF_{\alpha\beta}\left( z^\mu \epsilon^{\nu\rho\alpha\beta}+z^\rho \epsilon^{\mu\nu\alpha\beta}+z^\alpha\epsilon^{\mu\rho\nu\beta}+z^\beta \epsilon^{\mu\rho\alpha\nu}\right)z_\rho  \nn 
&=& -z^\mu F_{\mu\nu}W\tilde{F}^{\nu\rho}z_\rho -z^2 F^{\mu\nu}W\tilde{F}_{\mu\nu} - 2z_\rho \tilde{F}^{\rho\beta}WF_{\beta\alpha}z^\alpha.
\eeq
 The coordinates of $F,\tilde{F}$ are arbitrary.} 
\beq
2z^\nu\left( F_{\nu\mu} W\tilde{F}^{\mu\rho}+\tilde{F}_{\nu\mu} W F^{\mu\rho}\right)z_\rho = -z^2 F^{\mu\nu}W\tilde{F}_{\mu\nu}, \label{schouten}
\eeq
which follows from the Schouten identity. It is important to notice that this formula   eliminates the $1/z^2$ singularity without requiring the assumption of the symmetric limit (\ref{limit})  which, in the nonlocal case, might be generalized  as  \cite{Agaev:2014wna} 
\beq
z^\mu = \delta^\mu_- z^- + x^\mu, \qquad z^2=x^2, \qquad \lim_{x\to 0}\frac{x^\mu x^\nu}{x^2} \to \frac{g^{\mu\nu}}{4}. \label{zx}
\eeq
While the leading term (\ref{anom}) is not affected,\footnote{  Note that the  seemingly divergent expression  
\beq
\frac{(z^-)^2}{x^2}\left(\tilde{F}^{+\mu}F_{\mu}^{\ +}+F^{+\mu}\tilde{F}_\mu^{\ +}\right),
\eeq
identically vanishes. } 
this prescription induces spurious  contributions from expanding the field strength tensors around $x = 0$ such as  $z^-D^\alpha F_{\alpha\mu}\tilde{F}^{\mu+}$ 
(see the second equation of the right column on page 20  of 
\cite{Agaev:2014wna}). The present derivation based on the exact identity (\ref{schouten}) indicates that these extra operators are either     artifacts of the prescription (\ref{zx}), or should sum up to zero.  
Incidentally, by taking  the local limit $W\to 1$ in (\ref{schouten}), one finds a simpler identity  
\beq
4z^\nu F_{\nu\mu}\tilde{F}^{\mu\rho}z_\rho = -z^2 F^{\mu\nu}\tilde{F}_{\mu\nu}. \label{newid}
\eeq
This shows that the symmetric limit procedure (\ref{limit}) is actually unnecessary even in the textbook derivation of the chiral anomaly \cite{Peskin:1995ev}.  It is unclear to us whether this has been previously noted in the literature.

Using (\ref{anom}), we can write 
\beq
{\cal D}_\mu \left[\bar{\psi}(z_2)W\gamma^\mu \gamma_5 \psi(z_1)\right] &=&\left[O_F(z_2,z_1)\right]_{reg}-\frac{n_f\alpha_s}{2\pi}\int_{0}^{1} d\alpha \int_{0}^{1-\alpha} d\beta  F^{\mu\nu}(z_{12}^{\beta}) \tilde{W}\tilde{F}_{\mu\nu}\left(z_{21}^{\alpha}\right) .
\label{main}
 \eeq
The notation $[...]_{reg}$  means that the leading $1/z^2$ singularity  has been subtracted to avoid double counting. Although $[O_F]_{reg}$ still contains the logarithmic terms $\ln z^2$, it smoothly vanishes in the $z^\mu\to 0$ limit. In this way, the local chiral anomaly relation (\ref{ad2}) is recovered as $z^\mu\to 0$.

For the present purpose, we need to take the limit $z^2 \rightarrow 0$ in (\ref{main}) keeping $z^-$ finite.  
The issue here is precisely the difference in UV behaviors between composite operators separated by  light-like and non-light-like distances \cite{Balitsky:1987bk,Ji:2013dva, Radyushkin:2017cyf,Radyushkin:2019mye}, which has recently become a topic of great interest in lattice QCD. 
One utilizes factorization properties enabling a perturbative matching of space-like separated operators to light-like separated operators. The infrared sensitivity $1/\epsilon_{IR}+\ln \mu_{IR}^2$ in (\ref{exact}) cancels out in this procedure. 
Schematically, for a generic set of bi-local composite operators $A_i$, which mix under renormalization and whose component fields are separated by $z^{\mu}$, we can write
\begin{align}
A_i(z^2) = \sum_j  C_{ij}(\ln(-z^2 \mu_{UV}^2)) \otimes A_j(z^2 = 0, \mu_{UV}^2) + \mathcal O(z^2),
\label{matching}
\end{align}
where $\otimes$ denotes some convolution product that also includes mixing among the $A_i$. $\mu^2_{UV}$ is a factorization scale corresponding to the UV renormalization of the light-like operator $A_j(z^2 = 0)$ renormalized in dimensional regularization. $C_{ij} = \delta_{ij}  + \mathcal O(\alpha_s)$ is a perturbative matching kernel that depends on $z^2$ through polynomials in $\ln(-z^2\mu^2_{UV}  )$ at fixed orders in $\alpha_s$. 
Note that the factorization formula \eqref{matching} assumes that $A_i(z^2)$ diverges at most logarithmically as $z^2 \rightarrow 0$. For this reason, it was necessary to isolate the anomaly contribution from $O_F$. 
Applying \eqref{matching} to each of the operators shown in \eqref{main}, we arrive at 
\beq
{\cal D}_\mu \left[\bar{\psi}(z_2^-)W\gamma^\mu \gamma_5 \psi(z_1^-)\right] &=&O_F(z_2^-,z^-_1)-\frac{n_f\alpha_s}{2\pi}\int_{0}^{1} d\alpha \int_{0}^{1-\alpha} d\beta  F^{\mu\nu}(z_{12}^{\beta-}) \tilde{W}\tilde{F}_{\mu\nu}\left(z_{21}^{\alpha-}\right) +\cdots,
\label{main2}
 \eeq
where all the operators are now on the light-cone. They are renormalized at  $\mu_{UV}^2\sim -1/z^2$ in order to eliminate the large logarithms. 
The corrections $\cdots$ contain non-logarithmic  $\mathcal O(\alpha_s)$ corrections from the matching kernel $C$ in \eqref{matching} and higher twist corrections of order ${\cal O}(z^2) = {\cal O}(1/\mu_{UV}^2)$. In the following, we shall adopt  the leading logarithmic approximation and neglect these corrections.

 (\ref{main2}) was first derived in \cite{Mueller:1997zu} 
where the anomaly term $F\tilde{F}$ was added by hand by requiring  that both sides of (\ref{main2}) have the same matrix element between two on-shell gluon external states. This in particular means that $O_F$ evaluated exactly on the light-cone does not contain the anomaly.  (Hence the symbol $[...]_{reg}$ is unnecessary in (\ref{main2}).)  The details of this calculation were not given in \cite{Mueller:1997zu}. We have repeated this calculation and provide an  outline in Appendix A since we find it instructive. 
It should be mentioned, however, that two-gluon matrix elements are sensitive only to the Abelian part of the operator 
\beq
F^{\mu\nu}\tilde{F}_{\mu\nu}\sim \frac{1}{2}\epsilon_{\mu\nu\rho\lambda}(\partial^\mu A^\nu -\partial^\nu A^\mu)(\partial^\rho A^\lambda -\partial^\lambda A^\rho),  \label{abe}
\eeq
 and the non-Abelian part has to be supplemented by requiring gauge invariance or consistency with the chiral anomaly. Besides, the two-gluon calculations do not capture the all-order summation of gluons described by the Wilson line which is of course needed for  gauge invariance.  The same comments apply to the recent calculations \cite{Tarasov:2020cwl,Tarasov:2021yll,Bhattacharya:2022xxw,Bhattacharya:2023wvy}.  In the present derivation (see also  \cite{Agaev:2014wna}), the non-Abelian part and the Wilson line are fully included. 
 It also clarifies the accuracy (i.e., the leading logarithmic approximation) of the relation obtained. \\

Let us establish the connection to the work \cite{Bhattacharya:2022xxw,Bhattacharya:2023wvy}. 
First we need the precise definition of the  twist-four GPD mentioned in (\ref{4})    
\beq
\tilde{{\cal F}}(x,\xi,t)\equiv  \frac{iP^+}{M}\int \frac{dz^-}{2\pi} e^{ixP^+z^-} \langle p'|F^{\mu\nu}(-z^-/2)\tilde{W}\tilde{F}_{\mu\nu}(z^-/2)|p\rangle \, , \label{two}
\eeq
where $\Delta^\mu=p'^\mu-p^\mu$ and $P^\mu = \frac{1}{2}(p'^\mu+p^\mu)$.  $t=\Delta^2$ is the momentum transfer and  $\xi=-\Delta^+/(2P^+)$  is the skewness variable. We work in a frame where the transverse components of $P^\mu$ vanish $P^{i}=0$. 
 (Compared to the definition given in  \cite{Bhattacharya:2023wvy}, we have removed the spinor product $\bar{u}(p')\gamma_5u(p)$ in the denominator, which is convenient for the present purpose.)  We also introduce  a new GPD 
 \beq
 iO_F(x,\xi,t)=P^+\int \frac{dz^-}{2\pi}e^{ixP^+z^-}\langle p'|O_F(-z^-/2,z^-/2)|p\rangle, \label{ofgpd}
 \eeq
which has the property $\int dx O_F(x,\xi,t)=0$.

We now take the proton matrix element of (\ref{main}) and set $z_2=-\frac{z}{2}$ and $z_1=\frac{z}{2}$. On the right hand side, we identify the distribution (\ref{two})  
 \beq
&&\int_{0}^{1} d\alpha \int_{0}^{1-\alpha} d\beta  \langle p'|F^{\mu\nu}\left(\frac{2\beta-1}{2}z^-\right) \tilde{W}\tilde{F}_{\mu\nu}\left(\frac{1-2\alpha}{2} z^-\right)|p\rangle \nn
&& =-iM\int dx' \int_0^1d\alpha \int_0^{1-\alpha}d\beta e^{-i\xi P^+z^-(\beta-\alpha)} e^{-ix'P^+z^-(1-\alpha-\beta)} \tilde{{\cal F}}(x',\xi,t) \nn 
&& =\int dx' \frac{-iM\tilde{{\cal F}}(x',\xi,t)}{2\xi (\xi^2-x'^2)(P^+z^-)^2}\left[(x'+\xi)\left(e^{i\xi P^+z^-}-e^{-i\xi P^+z^-}\right)-2\xi\left(e^{i\xi P^+z^-}-e^{-ix'P^+z^-}\right)\right].
 \eeq
 On the left hand side, the translation derivative ${\cal D}_\mu$ is converted into  $i\Delta_\mu$. We thus find, after Fourier transforming in $z^-$, 
 \beq
 && i\Delta_\mu \int\frac{dz^-}{2\pi}e^{ixP^+z^-} \langle p'|\bar{\psi}(-z^-/2)\gamma^\mu \psi(z^-/2)|p\rangle
 \nn
 && =\frac{\alpha_s }{2\pi}\int dx'\frac{iM}{2\xi(\xi^2-x'^2)(P^+)^2}\int \frac{dz^-}{2\pi}\frac{1}{(z^-)^2}\nn
 && \qquad \times \left[(x+\xi)\left(e^{i(x+\xi) P^+z^-}-e^{i(x-\xi) P^+z^-}\right)-2\xi\left(e^{i(x+\xi) P^+z^-}-e^{i(x-x')P^+z^-}\right)\right]\tilde{{\cal F}}(x',\xi,t) +\frac{i}{P^+}O_F(x,\xi,t) \nn
 && = \frac{\alpha_s}{2\pi} \frac{iM}{4P^+}\int dx' \frac{-(\xi+x')|x-\xi|-(\xi-x')|x+\xi|+2\xi|x-x'|}{\xi(x'^2-\xi^2)} \tilde{{\cal F}}(x',\xi,t)+\frac{i}{P^+}O_F(x,\xi,t)  \nn 
 && = \frac{\alpha_s}{2\pi} \frac{iM}{P^+}\left[\int_x^1 dx' \frac{x'-x}{x'^2-\xi^2}-\theta(\xi-x)\int_{-1}^1 dx' \frac{(\xi+x')(\xi-x)}{2\xi (x'^2-\xi^2)}
\right]\tilde{{\cal F}}(x',\xi,t)+\frac{i}{P^+}O_F(x,\xi,t),
 \eeq
 where we assumed $x>0$. 
Using the property $\tilde{\cal F}(x,\xi,t)=\tilde{\cal F}(-x,\xi,t)$, the last line can be written as 
\beq
\frac{\alpha_s}{2\pi} \frac{iM}{P^+}\left[\int_x^1 dx' \frac{x'-x}{x'^2-\xi^2}-\theta(\xi-x)\int_0^1 dx' \frac{\xi-x}{ x'^2-\xi^2}
\right]\tilde{{\cal F}}(x',\xi,t)  \equiv \frac{\alpha_s}{2\pi} \frac{iM}{P^+}\tilde{C}^{anom}\otimes \tilde{{\cal F}}(x,\xi,t).
 \eeq
 For $x<0$, we have 
 \beq
 \tilde{C}^{anom}\otimes \tilde{\cal F}(x,\xi,t)\equiv\left[\int_{-1}^x dx' \frac{x-x'}{x'^2-\xi^2}-\theta(x+\xi)\int^0_{-1} dx' \frac{\xi+x}{ x'^2-\xi^2} 
\right]\tilde{{\cal F}}(x',\xi,t).
 \eeq
 Note the following property
 \beq
 \int_{-1}^1 dx \tilde{C}^{anom}\otimes \tilde{\cal F}(x,\xi,t) = \frac{1}{2}\int_{-1}^1 dx\tilde{\cal F}(x,\xi,t). 
 \eeq
The convolution kernel $\tilde{C}^{anom}$ exactly agrees with  the one encountered in \cite{Bhattacharya:2022xxw,Bhattacharya:2023wvy} in the calculation of the Compton amplitudes. (We have changed the normalization by a  factor of 4.) 
We thus arrive at 
\beq
\Delta_\mu P^+\int\frac{dz^-}{2\pi}e^{ixP^+z^-} \langle p'|\bar{\psi}(-z^-/2)W\gamma^\mu\gamma_5 \psi(z^-/2)|p\rangle = \frac{n_f\alpha_sM}{2\pi} \tilde{C}^{anom}\otimes \tilde{{\cal F}}(x,\xi,t) +O_F(x,\xi,t). \label{i}
\eeq

\section{Anomaly-mediated relations between GPDs}

In \cite{Mueller:1997zu}, the authors studied the relation  (\ref{main2}) in the context of polarized DIS. Since the divergence of the current is associated with nonzero momentum transfer ${\cal D}_\mu\sim \Delta_\mu$, it is more natural to discuss its implications on GPDs. 
We now use (\ref{i}) to derive new identities among GPDs. 
On the left hand side, we recognize the standard definitions of the twist-two $\gamma^{\mu=+}$, twist-three $\gamma^{\mu=i}$ and twist-four $\gamma^{\mu=-}$ GPDs. Explicitly,  
\beq
&&\Delta^-P^+\int \frac{dz^-}{2\pi} e^{ixP^+z^-}\langle p'|\bar{\psi}(-z/2)\gamma^+ \gamma_5\psi(z/2)|p\rangle = \Delta^-\bar{u}(p') \left[\tilde{H}\gamma^+\gamma_5+\tilde{E}\frac{\Delta^+}{2M}\gamma_5\right]u(p), \label{234}\\
 &&\Delta_i P^+\int \frac{dz^-}{2\pi} e^{ixP^+z^-}\langle p'|\bar{\psi}(-z/2)\gamma^i \gamma_5\psi(z/2)|p\rangle \nn
 &&\qquad = \Delta_i \bar{u}(p') \left[\tilde{H}_3\gamma^i\gamma_5+\tilde{E}_3\frac{\Delta^i}{2M}\gamma_5 
+\tilde{G}_{3}\frac{\Delta^i}{P^+}\gamma^+\gamma_5+i\tilde{G}'_{3}\epsilon^{ij}\frac{\Delta_j}{P^+}\gamma^+ \right]u(p), \nn
&& \Delta^+P^+\int \frac{dz^-}{2\pi} e^{ixP^+z^-}\langle p'|\bar{\psi}(-z/2)\gamma^- \gamma_5\psi(z/2)|p\rangle =\Delta^+\bar{u}(p') \left[\tilde{H}_4\gamma^-\gamma_5+\tilde{E}_4\frac{\Delta^-}{2M}\gamma_5\right]u(p) . \nonumber
\eeq
All the GPDs are functions of $x,\xi$ and $t$ (and the renormalization scale), $\tilde{H}=\tilde{H}(x,\xi,t)$, etc. Also, the summation over quark flavors is implied, $\tilde{H}=\sum_q \tilde{H}_q$, etc. 
The twist-3 GPDs are from \cite{Kiptily:2002nx} where we redefined $\tilde{H}+\tilde{G}_2\to \tilde{H}_3$ and $\tilde{E}+\tilde{G}_1\to \tilde{E}_3$. (We also redefined $\tilde{G}_4\to \tilde{G}'_3$.)  
The twist-4 GPDs are parametrized differently from  
\cite{Meissner:2009ww} but the two parametrizations  are equivalent in the present frame $P^i=0$.  
On the right hand side, the twist-4 pseudoscalar GPD (\ref{two}) is parametrized as  \cite{Hatta:2020ltd}
\beq
\tilde{{\cal F}}(x,\xi,t)=\frac{1}{M}\bar{u}(p')\left[ \Delta^-\gamma^+\gamma_5 \tilde{\cal F}_2+\Delta_i \gamma^i\gamma_5 \tilde{\cal F}_3+\Delta^+\gamma^-\gamma_5 \tilde{\cal F}_4\right]u(p). \label{fpara}
\eeq
Lorentz invariance requires that 
\beq
\int dx \tilde{\cal F}_2(x,\xi,t)=\int dx \tilde{\cal F}_3=\int dx \tilde{\cal F}_4 = \frac{i}{2M}\frac{\langle p'|F^{\mu\nu}\tilde{F}_{\mu\nu}|p\rangle}{\bar{u}(p')\gamma_5u(p)}. 
\eeq
Both the numerator and the denominator vanish in the forward limit $\Delta^\mu\to 0$, but the ratio, which is a function only of $t$,  has a finite limit as $t\to 0$.  In fact, it is proportional to the quark helicity contribution to the proton spin 
\beq
\lim_{t\to 0} \frac{i}{2M}\frac{\langle p'|\frac{n_f\alpha_s}{4\pi}F^{\mu\nu}\tilde{F}_{\mu\nu}|p\rangle}{\bar{u}(p')\gamma_5u(p)} = \Delta \Sigma. \label{sigma}
\eeq
A partial insight into the distribution $\tilde{\cal F}_{2,3,4}$ can be obtained from the equation of motion relation  \cite{Hatta:2020ltd} 
\beq
\tilde{\cal F}(x,\xi,t)&=& \frac{i\Delta_\mu}{xM} \int \frac{dz^-}{2\pi}e^{ixP^+z^-} \langle p'| \tilde{F}^\mu_{\ \nu}(-z^-/2)F^{\nu +}(z^-/2)-F^{\nu+}(-z^-/2)\tilde{F}^{\mu}_{\ \nu}(z^-/2)|p\rangle \nn
&& -\frac{i}{xM}\int \frac{dz^-}{2\pi}e^{ixP^+z^-}\int_{-z^-/2}^{z^-/2}dw^- \nn  && \qquad \times \langle p'|\tilde{F}_{\mu\nu}(-z^-/2)gF^{+\mu}(w^-)F^{+\nu}(z^-/2)-F^{+\nu}(-z^-/2)F^{+\mu}(w^-)\tilde{F}_{\mu\nu}(z^-/2)|p\rangle. \label{19}
\eeq 
Comparing with (\ref{fpara}), we find, in the forward limit, 
\beq
\tilde{\cal F}_2(x) = -\Delta G(x) + ({\rm twist \ four}), \label{deltag}
\eeq
where $\Delta G(x)$ is the polarized gluon distribution. Similarly, $\tilde{\cal F}_3(x)$ is related to the twist-three distribution ${\cal G}_{3T}(x)$ (the gluonic counterpart of the $g_T(x)$ distribution) for the transversely polarized proton  \cite{Ji:1992eu,Hatta:2012jm}. 
The twist-four GPD $O_F$ (\ref{ofgpd}) also admits a similar decomposition 
\beq
O_F(x,\xi,t)= \bar{u}(p') \left[\Delta^-\gamma^+\gamma_5O_{F2}+\Delta_i \gamma^i\gamma_5O_{F3}+\Delta^+\gamma^-\gamma_5O_{F4}\right]u(p),
\eeq
with 
\beq
\int dx O_{F2}(x,\xi,t)=\int dx O_{F3}=\int dx O_{F4}=0.  \label{zero}
\eeq

Let us assume that $\Delta^+=\xi=0$. Then 
\beq
\Delta^-= \xi\frac{M^2+\frac{\Delta_i\Delta^i}{4}}{P^+(1-\xi^2)}=0,
\eeq
and $t=\Delta_i\Delta^i$. From parity and time-reversal ($PT$) symmetry, it is easy to see that  $\tilde{G}_{3}$ is odd under $\xi \to -\xi$, hence it vanishes when $\xi=0$. (\ref{i}) reduces to 
\beq
\tilde{H}_3+\frac{t}{4M^2}\tilde{E}_3 = \frac{n_f\alpha_s}{2\pi} \tilde{C}^{anom}\otimes \tilde{\cal F}_3 +O_{F3}, \label{res1}
\eeq
where we used   $\Delta_i\bar{u}\gamma^i\gamma_5 u =\bar{u}\Slash \Delta \gamma_5 u = 2M\bar{u}\gamma_5 u$. This result is valid for $\xi=0$ but arbitrary $t$.  
Integrating over $x$, we exactly reproduce the relation (\ref{form1}) among the form factors due to (\ref{zero}) and the relations such as 
\beq
\int_{-1}^1 dx \tilde{H}_3(x,\xi,t)=g_A(t), \qquad \int_{-1}^1 dx \tilde{E}_3(x,\xi,t)=g_P(t).
\eeq

Next consider the case $\Delta_i=0$ where 
\beq
t=2\Delta^+\Delta^-=-\frac{4\xi^2}{1-\xi^2}M^2.
\eeq
Using  the following identities 
\beq
&&\Delta^+\bar{u}(p')\gamma^-\gamma_5 u(p) = \Delta^-\bar{u}(p')\gamma^+ \gamma_5u(p), \nn 
&& 2M\bar{u}(p')\gamma_5u(p)=\bar{u}(p')\Slash \Delta \gamma_5u(p) =2\Delta^-\bar{u}(p')\gamma^+\gamma_5u(p),
\eeq
we obtain 
\beq
\tilde{H}+\tilde{H}_4 +\frac{t}{4M^2}(\tilde{E}+\tilde{E}_4)=\frac{n_f\alpha_s}{2\pi}\tilde{C}^{anom}\otimes (\tilde{\cal F}_2+\tilde{\cal F}_4) + O_{F2}+O_{F4}.
\label{res2}
\eeq
Integrating over $x$, we again recover (\ref{form1}) because 
\beq 
\int dx \tilde{E}=\int dx\tilde{E}_4=g_P(t), \quad \int dx \tilde{H}=\int dx\tilde{H}_4=g_A(t).
\eeq

In order to exhibit the `anomaly poles' 
mentioned in the introduction, we just need to rewrite (\ref{res1}) and (\ref{res2}) as 
\beq
\tilde{E}_3 && =\frac{4M^2}{t}\left( \frac{n_f\alpha_s}{2\pi}\tilde{C}^{anom}\otimes \tilde{\cal F}_3 -\tilde{H}_3 +O_{F3}\right) \qquad (\xi=0), \nn
\tilde{E}+\tilde{E}_4 && =\frac{4M^2}{t} \left(\frac{n_f\alpha_s}{2\pi}\tilde{C}^{anom}\otimes (\tilde{\cal F}_2+\tilde{\cal F}_4)-\tilde{H}-\tilde{H}_4 +O_{F2}+O_{F4}\right) \qquad \left( \xi^2=\frac{-t}{4M^2-t} 
\right) . \label{apole}
\eeq
 In both equations, the limit $t\to 0$ is well-defined, and  none of the distributions in (\ref{apole}) is expected to vanish in this limit. In particular, $\tilde{\cal F}$'s are finite at $t=0$ since their integrals (\ref{sigma}) are finite.  Therefore, the first term on the right hand side diverges as $t\to 0$. 
 This is the anomaly pole  that has been identified in the perturbative calculations of the Compton amplitude and the polarized GPD   \cite{Bhattacharya:2022xxw,Bhattacharya:2023wvy} (and previously in the context of polarized DIS \cite{Tarasov:2020cwl,Tarasov:2021yll}),  although these leading-twist calculations are blind to the difference between $\tilde{E}$ and $\tilde{E}_{3,4}$, etc.  We emphasize that all the distributions in (\ref{apole}) as well as the  kinematical variable $t$ are defined at the hadronic (not partonic) level. 
 
As already mentioned, the anomaly pole cancels another massless pole due to the exchange of the primordial $\eta_0$ meson in the limit $t\to 0$. The latter must be present because if it were not for the chiral  anomaly,\footnote{This happens, for example, in the large-$N_c$ limit of QCD. The anomaly term vanishes because it is proportional to $\alpha_s =\frac{g^2}{4\pi} = \frac{\lambda}{4\pi N_c}\to 0$ where $\lambda=g^2N_c$ is the 't Hooft coupling. The $\eta'$ meson mass goes to zero in this limit. } 
the GPD $\tilde{E}$ would have a massless pole due to the $\eta_0$ meson exchange. However, in contrast to the anomaly pole, the $\eta_0$ pole does not show up in perturbation theory, and has been  treated in an effective action approach \cite{Tarasov:2021yll}, or  introduced by hand \cite{Bhattacharya:2022xxw,Bhattacharya:2023wvy}.   We have now  explicitly derived the pole and found  that the residue consists of the $\tilde{H}$-type GPDs and the twist-four GPDs $O_F$.  As a result of this cancellation, the pole of  $\tilde{E}$ is shifted to the physical $\eta'$ meson mass $t=m_{\eta'}^2$. This is the GPD version of the well-known scenario in hadron physics \cite{Witten:1979vv,Veneziano:1979ec}. 

 For phenomenological purposes, one may implement various approximations. If one  ignores  the differences due to different twists, namely  $\tilde{H}_{3,4}\approx \tilde{H}$, $\tilde{\cal F}_{3,4}\approx \tilde{\cal F}_2$, etc., from (\ref{i})  one immediately obtains    
 \beq
 \tilde{E}(x,\xi,t) \approx \frac{4M^2}{t} \left(\frac{n_f\alpha_s}{2\pi}\tilde{C}^{anom}\otimes \tilde{\cal F}_2-\tilde{H}+O_{F2}\right). \label{last}
 \eeq
Within this approximation,  there is no restriction on $\xi$ and $t$. 
In the forward limit
\beq
O_{F2}(x)&\approx & \sum_q\Delta q(x) - \frac{n_f\alpha_s}{2\pi} \int_x^1 \frac{dx'}{x'} \left(1-\frac{x}{x'}\right)\tilde{\cal F}_2(x')
\nn &=&\sum_q\Delta q(x) + \frac{n_f\alpha_s}{2\pi} \int_x^1 \frac{dx'}{x'} \left(1-\frac{x}{x'}\right)\Delta G(x') +({\rm twist\  four}),
\label{ofalt} 
\eeq
where $\Delta q(x)$ is the polarized quark distribution and  we used (\ref{deltag}). 
The first moment reads
\beq
\int dx O_{F2}(x) = \Delta \Sigma + \frac{n_f\alpha_s}{2\pi}\Delta G +\cdots. \label{of}
\eeq
It is well known that this linear combination is renormalization group invariant to all orders  \cite{Altarelli:1990jp}. The omitted term in (\ref{of}) derives from the three-gluon, twist-four correlator $\langle \tilde{F}_{\mu\nu}F^{+\mu}F^{+\nu}\rangle$ in   (\ref{19}), and this cancels the linear combination \cite{Hatta:2020ltd} (see also \cite{Balitsky:1991te}), ensuring the property $\int dxO_{F2}(x)=0$. 
As a further approximation, it is  tempting to ignore $O_F$ altogether. (\ref{ofalt}) then implies a specific relation between $\Delta q(x)$ and $\tilde{\cal F}_2(x)$ at $t=0$ 
\beq
\sum_q\Delta q(x) \approx \frac{n_f\alpha_s}{2\pi}\int_x^1 \frac{dx'}{x'}\left(1-\frac{x}{x'}\right)\tilde{\cal F}_2(x'). \label{deltaqapp}
\eeq
This is similar to the relation obtained for the $g_1$ structure function  $g_1(x)=\frac{1}{2}\sum_q e_q^2 (\Delta q(x)+\Delta \bar{q}(x))+\cdots$ \cite{Tarasov:2021yll} using an effective theory of the $\eta_0$ meson. While the $x$-integral of (\ref{deltaqapp}) gives an identity $\Delta\Sigma=\Delta \Sigma$, in general the  $O_F$ term is needed  beyond the first moment.

Finally, let us comment on the effect of the current quark mass $m_q$ which has been neglected in the above calculation as well as in the recent works  \cite{Tarasov:2020cwl,Tarasov:2021yll,Bhattacharya:2022xxw,Bhattacharya:2023wvy}. In the perturbative calculation of the triangle \cite{Dolgov:1971ri,Horejsi:1985qu,Coriano:2014gja,Coriano:2023gxa,Coriano:2024ive} and box \cite{Castelli:2024eza} diagrams,  a finite fermion mass $m$ eliminates the anomaly pole at $t=0$ and converts it into a branch cut in the timelike region starting at $t= 4m^2$.  While this observation may be relevant to theories like QED, in QCD a  cut in the region $t\ge 4m_q^2$, representing a physical quark-antiquark threshold,  does not exist in hadronic form factors and GPDs because of confinement. In the case of the nucleon isovector axial form factor,   it is well known \cite{Nambu:1960xd} that the effect of the quark mass boils down to a slight shift of the pole position from $t=0$ to the physical pion mass 
\beq
g^{(3)}_P(t) \sim  \frac{1}{t}\to \frac{1}{t-m_\pi^2}.
\eeq
There is still a pole, not a cut, and $m_\pi^2$ is linearly proportional to the quark mass $m_\pi^2\propto m_q$. 
 The same pion pole is expected also in the corresponding GPD $\tilde{E}^{(3)}$ \cite{Penttinen:1999th}. 
In the case of the singlet axial form factor, one expects instead \cite{Veneziano:1979ec}
\beq
g_P(t) \sim \frac{1}{t}\to \frac{1}{t-m_\pi^2} \to \frac{1}{t-m_\pi^2+\frac{4n_f}{f^2_{\eta'}}\chi} = \frac{1}{t-m_{\eta'}^2},
\eeq
where $f_{\eta'}$ is the decay constant of the $\eta'$ meson and $\chi$ is the  topological susceptibility of the pure Yang-Mills theory
 \beq
 \chi= i\int d^4x \langle 0|{\rm T}\left\{\frac{\alpha_s}{8\pi}F\tilde{F}(x)\frac{\alpha_s}{8\pi}F\tilde{F}(0)\right\}|0\rangle.
 \eeq
 In the first step, the primordial $\eta_0$ meson gets a small mass equal to the pions' due to the  finite quark mass. Subsequently, it acquires a large mass 
 via the standard large-$N_c$ argument \cite{Witten:1979vv,Veneziano:1979ec} (or the pole cancellation, in our language).  We expect that, at finite quark mass, a similar shift $t\to t-m_\pi^2$ (or its $x$-dependent generalization)  occurs in the anomaly pole of  $\tilde{E}$ (\ref{apole}) due to the twist-three GPD 
\beq
2m_qP^+\int \frac{dz^-}{2\pi} e^{ixP^+z^-} \langle p'|\bar{\psi}(-z^-/2)W\gamma_5\psi(z^-/2)|p\rangle,
\eeq
to be added to the right hand side of (\ref{i}). 
However, the precise nature of this mechanism is nonperturbative and cannot be derived from perturbation theory involving massive quarks.

\section{Conclusions}

In this paper,  we have revisited and improved the derivation of the nonlocal chiral anomaly in \cite{Mueller:1997zu,Agaev:2014wna} by  clarifying the subtleties involved. Based on this, we have  derived  relations among the $\tilde{H}$-type and $\tilde{E}$-type polarized quark GPDs mediated by the chiral anomaly. The result can be used to establish  the nonperturbative foundation of the anomaly pole $1/t$  previously identified in perturbation theory \cite{Tarasov:2020cwl,Tarasov:2021yll,Bhattacharya:2022xxw,Bhattacharya:2023wvy}.  We have consolidated the conclusion in \cite{Bhattacharya:2022xxw,Bhattacharya:2023wvy} that the pole belongs to the GPD $\tilde{E}$, and gets canceled by another massless pole which can be interpreted as the $\eta_0$ meson exchange. This scenario, originally envisaged within the context of polarized DIS  \cite{Jaffe:1989jz,Tarasov:2020cwl,Tarasov:2021yll}, has now been explicitly demonstrated via the newly derived  relations between GPDs of various twists. Our work offers novel  insights into the nonperturbative structure of GPDs which are consistent with the wisdom of  hadron physics \cite{Witten:1979vv,Veneziano:1979ec,Shore:2007yn}. We hope this further motivates the study of GPDs at future experimental facilities such as SoLID at Jefferson Laboratory  \cite{JeffersonLabSoLID:2022iod} and the Electron-Ion Collider at the Brookhaven National Laboratory \cite{AbdulKhalek:2021gbh}.

We have entirely omitted the discussion of the trace anomaly. A perturbative analysis of the Compton amplitudes \cite{Bhattacharya:2022xxw,Bhattacharya:2023wvy} indicates that a $1/t$  pole is induced  in the unpolarized GPDs $H$ and $E$. This can be interpreted as the nonlocal version of the trace anomaly pole  \cite{Giannotti:2008cv,Armillis:2010qk}. Our result for the chiral anomaly strongly suggests the existence of  nonperturbative relations between  $H$, $E$ and the `gluon condensate' GPD \cite{Hatta:2020iin} mediated by the trace anomaly. Such relations will also provide new perspectives on the moments of the GPDs, namely the gravitational form factors.  Preliminary discussions were given in \cite{Bhattacharya:2022xxw,Bhattacharya:2023wvy}, but a rigorous analysis similar to the one we have been able to present here appears to be challenging. We leave this problem to future work. 

\section*{Acknowledgements}

We are grateful to Werner Vogelsang for the collaboration in \cite{Bhattacharya:2022xxw,Bhattacharya:2023wvy} which led to the present work and for his continuing support. We also 
thank Ian Balitsky, Vladimir Braun, Claudio Coriano and Raju Venugopalan for discussions and correspondence. 
The work of S.~B. has been supported by the Laboratory Directed Research and Development (LDRD) program of Los Alamos National Laboratory under project number 20240738PRD1.
S.~B. has also received support from the U.~S. Department of Energy through the Los Alamos National Laboratory. Los Alamos National Laboratory is operated by Triad National Security, LLC, for the National Nuclear Security Administration of U.~S. Department of Energy (Contract No. 89233218CNA000001).
Y.~H. and J.~S. were supported by the U.S. Department of Energy under Contract No. DE-SC0012704, and also by LDRD funds from Brookhaven Science Associates.

\appendix
\section{Two-gluon matrix elements}

In this Appendix, we compute the matrix elements 
\beq
\langle p'\epsilon'|{\cal D}_\mu \left[\bar{\psi}(z_2)W\gamma^\mu \gamma_5 \psi(z_1)\right]|p\epsilon\rangle, \qquad \langle p'\epsilon'|O_F(z_2,z_1)|p\epsilon\rangle, 
\eeq
to order $\alpha_s$ 
 between on-shell gluon states with $p^2=p'^2=0$ and $p\cdot \epsilon=p'\cdot \epsilon'=0$. We work in the gauge $z\cdot A=0$ ($z=z_1-z_2$) which implies $z\cdot \epsilon=z\cdot \epsilon'=0$. Let us first assume that $z^2\neq 0$. In this case, there is no divergence and we may use the usual four-dimensional $\gamma_5$. A straightforward calculation gives  
\beq
&&\langle p'\epsilon'|{\cal D}_\mu \left[\bar{\psi}(z_2)W\gamma^\mu \gamma_5 \psi(z_1)\right]|p\epsilon\rangle \nn 
&&=i\Delta_\mu \langle p'\epsilon'| \bar{\psi}(z_2)W\gamma^\mu \gamma_5 \psi(z_1)    |p\epsilon\rangle \nn &&= -2g^2n_f  \epsilon_{\alpha\beta\mu\lambda} \epsilon'^\mu \epsilon^\lambda(e^{ip'\cdot z_2}-e^{ip'\cdot z_1})(e^{-ip\cdot z_2}-e^{-ip\cdot z_1}) \left(\frac{p^\beta}{p\cdot z} - \frac{p'^\beta}{p'\cdot z} \right) \int \frac{d^dk}{(2\pi)^d}\frac{k^\alpha}{(k^2)^2}e^{ik\cdot z}. \label{2glu}
\eeq
 The result for $\langle p'\epsilon'|O_F(z_2,z_1)|p\epsilon\rangle$ is identical  as expected from the equation of motion relation (\ref{nonlocal}).  The $k$-integral gives ($d=4$)
\beq
 \int \frac{d^dk}{(2\pi)^d}\frac{k^\alpha}{(k^2)^2}e^{ik\cdot z} = -\frac{z^\alpha}{8\pi^2 z^2}.\label{kint}
 \eeq
Using the Schouten identity, one can check that 
\beq
\epsilon_{\alpha\beta\mu\lambda}\epsilon'^\mu \epsilon^\lambda(p'\cdot z p^\beta -p\cdot z p'^\beta)z^\alpha = \epsilon_{\rho\beta\mu\lambda}\epsilon'^\mu \epsilon^\lambda p'^\rho p^\beta z^2. 
\eeq
 Thus the singularity $1/z^2$ is canceled  and we find 
 \beq
&&\langle p'\epsilon'|{\cal D}_\mu \left[\bar{\psi}(z_2)W\gamma^\mu \gamma_5 \psi(z_1)\right]|p\epsilon\rangle =\langle p'\epsilon'|O_F(z_2,z_1)|p\epsilon\rangle\nn 
&& =  \frac{\alpha_sn_f}{\pi}\frac{1}{p\cdot z p'\cdot z} \epsilon_{\rho\beta\mu\lambda}\epsilon'^\mu \epsilon^\lambda p'^\rho p^\beta (e^{ip'\cdot z_2}-e^{ip'\cdot z_1})(e^{-ip\cdot z_2}-e^{-ip\cdot z_1}). \label{fin}
 \eeq
This agrees with the matrix element of the anomaly operator (\ref{anom}). 

We now consider the light-cone limit $z^\mu \to \delta^\mu_-z^-$ which is subtle. 
 Exactly on the light-cone, the integral (\ref{kint}) is UV divergent and the prescription for the $\gamma_5$ becomes an issue.  In the 't Hooft-Veltman-Breitenlohner-Maison (HVBM) scheme \cite{tHooft:1972tcz,Breitenlohner:1977hr}, the $\gamma_5$ in $d$-dimensions does not anti-commute with the $d-4$ dimensional components of the loop momentum $\Slash k$, and this generates an extra ${\cal O}(\epsilon)$ term in the computation of $\langle {\cal D}_\mu \left[\bar{\psi}(z_2^-)\gamma^\mu \gamma_5 \psi(z_1^-)\right]\rangle$. If and only if one is on  the light-cone, this term leaves a finite contribution because it gets multiplied by a UV-divergent $k$-integral.     The calculation is analogous to the local case \cite{Peskin:1995ev} but slightly more lengthy. We find 
\beq
\langle p'\epsilon'|{\cal D}_\mu \left[\bar{\psi}(z_2^-)W\gamma^\mu \gamma_5 \psi(z_1^-)\right]|p\epsilon\rangle =({\rm A2}) +\frac{\alpha_sn_f}{\pi}\frac{\epsilon_{\rho\beta\mu\lambda}\epsilon'^\mu \epsilon^\lambda p'^\rho p^\beta}{p^+p'^+(z^-)^2}  (e^{ip'^+ z^-_2}-e^{ip'^+ z^-_1})(e^{-ip^+ z^-_2}-e^{-ip^+ z^-_1}).
\eeq
 The added term exactly agrees with the light-cone limit of (\ref{fin}). On the other hand, $\langle O_F(z_2^-,z^-_1)\rangle$ is not affected by this subtlety and is still  given by (\ref{2glu}).

\bibliography{ref}

\end{document}